# Development and evaluation of 10-inch Photo-Multiplier Tubes for the Double Chooz experiment

T. Matsubara[a], T. Haruna[b,1], T. Konno[a], Y. Endo[b,2], M. Bongrand[c,3], H. Furuta[c], T. Hara[d], M. Ishitsuka[a],
T. Kawasaki[e], M. Kuze[a], J. Maeda[b], Y. Mishina[e], Y. Miyamoto[e], H. Miyata[e], Y. Nagasaka[f], Y. Sakamoto[g],
F. Sato[b], A. Shigemori[e], F. Suekane[c], T. Sumiyoshi[b], H. Tabata[c], N. Tamura[e]

[a]*Department of Physics, Tokyo Institute of Technology, Tokyo 152-8551, Japan.*
[b]*Department of Physics, Tokyo Metropolitan University, Tokyo 192-0397, Japan.*
[c]*Department of Physics, Tohoku University, Miyagi 980-8578, Japan.*
[d]*Department of Physics, Kobe University, Hyogo 657-8501, Japan.*
[e]*Department of Physics, Niigata University, Niigata 950-2181, Japan.*
[f]*Department of Computer Science, Hiroshima Institute of Technology, Hiroshima 731-5193, Japan.*
[g]*Department of Information Science, Tohoku Gakuin University, Miyagi 981-3193, Japan.*
[1]*Now at Canon Inc..*
[2]*Now at Nihon Koden.*
[3]*Now at Charge de Recherche, LAL - Universite Paris Sud 11.*

## Abstract

The goal of Double Chooz experiment is a precise measurement of the last unknown mixing angle $\theta_{13}$ using two identical detectors placed at far and near sites from Chooz reactor cores. The detector is optimized for reactor-neutrino detection using specially developed 10-inch PMTs. We developed two types of measurement systems and evaluated 400 PMTs before the installation. Those PMTs fulfill our requirements, and a half of those have been installed to the far detector in 2009. The character and performance data of the PMTs are stored in a database and will be referenced in analysis and MC simulation.

*Key words:*

## 1. Introduction

The Double Chooz experiment [1] is a neutrino oscillation experiment at Chooz nuclear power plant in France, which aims to accurately measure the last unknown mixing angle $\theta_{13}$, one of the most demanded parameters in neutrino physics. The current upper limit is $\sin^2 2\theta_{13} < 0.15$ (90 % C.L., for $\Delta m_{13}^2 = 2.5 \times 10^{-3}$ eV$^2$). We will measure the $\sin^2 2\theta_{13}$ with 5 times better sensitivity than the current limit measured by the previous CHOOZ experiment. Double Chooz will place two detectors of same structure, i.e. far and near detectors, to accomplish the systematic-free measurement of the neutrino disappearance. The experiment will start in year 2010 with the far detector only. The near detector will follow 1.5 years later.

Figure 1 shows a schematic view of the Double Chooz detector. Target and $\gamma$-catcher vessels are built with acrylic material. Target region is filled with gadolinium (Gd) doped liquid scintillator. Anti-electron neutrinos coming from reactor cores are detected through inverse $\beta$-decay process, $\bar{\nu}_e + p \to e^+ + n$, which generates a positron and a neutron. The positron deposits its energy in the liquid scintillator, then annihilates with electron. The spectrum of positron signal (prompt signal) ranges between 1 to 8 MeV. The Gd captures neutron and generates $\gamma$-rays with a total energy of about 8 MeV, giving a delayed signal. The $\gamma$-cacher region is filled with un-doped liquid scintillator in order to detect $\gamma$-rays escaped from target region. Buffer tank made of stainless steel surrounds the $\gamma$-catcher region. Buffer region is filled with non-scintillating mineral oil so that it reduces environmental $\gamma$-rays coming from Photo-Multiplier Tube (PMT) glass, for example. A total of 390 10-inch PMTs are mounted at the wall of buffer tank. Inner veto tank is filled with liquid scintillator and surrounded by steel shield. The scintillation light generated in inner-veto region is monitored by another type of 8-inch 78 PMTs, which are not described in this paper.

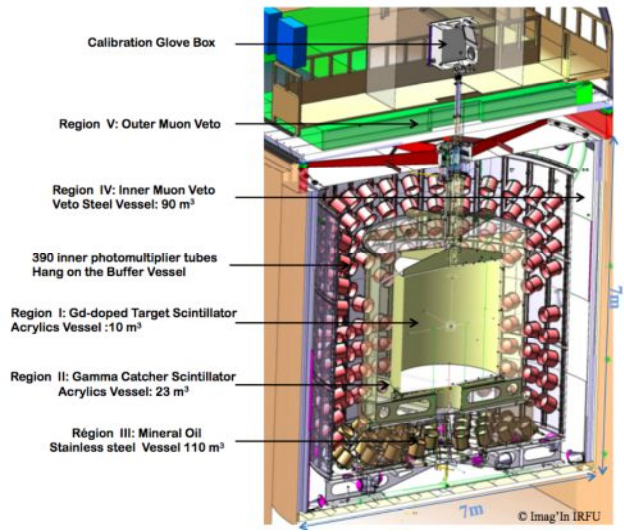

Figure 1: Double Chooz detector structure.

Double Chooz uses about 800 PMTs for the buffer region in two detectors. Japanese group is in charge of 400 PMTs, and the other half in Germany. This paper deals with the Japanese responsibility. They have been delivered to Tokyo Institute of Technology, tested and evaluated in detail one by one. Then they were transported to Max Planck Institute for nuclear physics (MPI-K) in Germany and tested again in order to check the damage during transportation. Those PMTs which satisfied the quality cut were assembled with support structure and a magnetic shield of $\mu$-metal [2] at MPI-K. Finally, those were randomly separated for near and far detectors to suppress systematic bias induced by personalities of each PMT.

This paper describes the PMT system for the Double Chooz experiment, PMT evaluation systems developed in Japan and results of the evaluation.

## 2. PMT system for the Double Chooz experiment

PMT system for the Double Chooz detector consists of PMT, magnetic shield made from $\mu$-metal, support structure, High Voltage (HV) supply, HV/signal splitter and cable down to front-end electronic (FEE) as shown in Figure 2. The HV system consists of HV crate and modules, named SY1527LC and A1535P made by CAEN S.p.A.. The FEEs, electronics originally developed by Drexel university for amplifying and shaping of PMT output, send signals to a trigger board and Flash ADC.

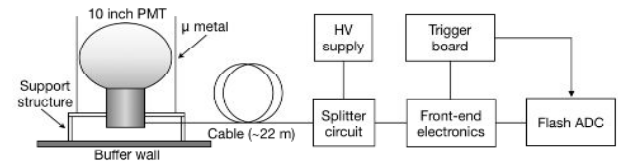

Figure 2: Schematic view of PMT system.

The splitter circuit is needed to extract signal from PMT since each PMT has a single cable, carrying HV and PMT signal in order to simplify the cabling work and to avoid ground-loop effects. It reduces the cost as well. Figure 3 shows the splitter circuit, which was developed by CIEMAT[4] group. High and low-pass filters and a current monitoring LED are implemented for each channel.

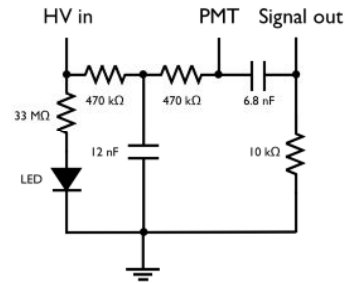

Figure 3: The splitter circuit.

### 2.1. PMT specification

The Double Chooz experiment employs the PMT developed by Hamamatsu Photonics K.K. (HPK) [3] based on R7081 PMT in order to achieve very low-background condition. The basic specifications and design of Hamamatsu R7081 are summarized in Table 1 and Figure 4.

The PMT glass was formed in platinum coating glass furnace, and the radioactive contamination of $^{238}$U, $^{232}$Th and $^{40}$K from the wall of the glass furnace was largely reduced, resulting in 13 ppb, 61 ppb and 3.3 ppb, respectively, as a median in glass of a PMT [4]. This dependence of the PMT response.

---
[4]Centro de Investigaciones Energeticas, Medioambientales y Tecnologicas, Spain

| Item | Specification |
|---|---|
| Wave length region | 300 nm~650 nm |
| Photo cathode | Bialkali |
| Peak wavelength | 420 nm |
| Diameter | $\phi$253 mm |
| Number of dynodes | 10 |
| Glass weight | ~1,150 g |

Table 1: Basic specification of R7081 [3]

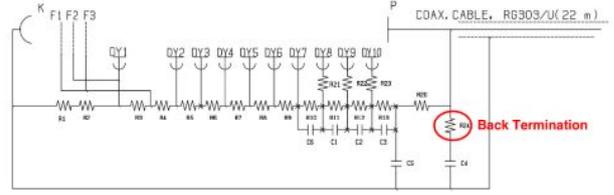

Figure 5: The bleeder circuit for R7081. R1, R2: 2.5 M$\Omega$, R3: 180 k$\Omega$, R4: 1.02 M$\Omega$, R5: 1.5 M$\Omega$, R6: 1 M$\Omega$, R7: 499 k$\Omega$, R8: 300 k$\Omega$, R9: 360 k$\Omega$, R10: 430 k$\Omega$, R11: 680 k$\Omega$, R12: 910 k$\Omega$, R13: 750 k$\Omega$, R20: 10 k$\Omega$, R21-R23: 100 $\Omega$, R24: 49.9 $\Omega$, C1-C3, C6: 10 nF, C4, C5: 4700 pF.

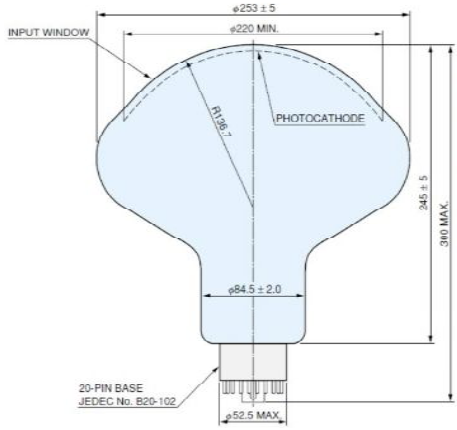

Figure 4: Design of Hamamatsu R7081.

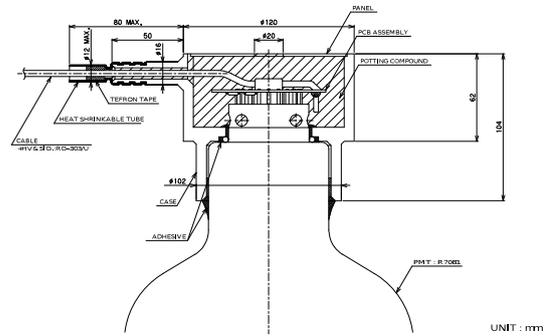

Figure 6: Design of PMT base part.

feature is especially important for the Double Chooz experiment since the oscillation maximum is at a few MeV and overlaps with radioactivities.

Dynode of R7081 has a box-line type structure. Figure 5 shows the bleeder circuit for Double Chooz PMT. R24 (49.9 $\Omega$) is called "back termination". It reduces the amplitude of the signal in half, but it quickly damps reflections of the signals caused by impedance miss-matching, which is an important feature to quickly recover from very large muon signals. The dynode circuit is also optimized to cover several hundred P.E.s with pulse linearity and to reduce ringing.

The PMT base part is molded by an epoxy glue. The purpose of this molding is to protect electric parts and soldering parts from possible degradation due to oil, as well as to prevent contamination in oil by possible solution from the bleeder material and soldering. This oil-proof base was developed for the KamLAND experiment and has been reliably used in paraffin oil for more than 6 years. Figure 6 shows a part of assembled base circuit. The cable outlet direction from the PMT is set to random in units of 45 degrees with respect to the dynode orientation in order to average the azimuthal angle

### 2.2. Requirements for PMT performance

Several product-quality standards were set to HPK as shown in Table 2. The standards were also used as reference for our evaluation result in order to select candidate PMTs for installation. Those which show relatively low performance were kept as backups.

Figure 7 shows the charge distribution obtained by shining a PMT by laser pulses with an intensity corresponding to give 0.25 photoelectron (P.E.) of expected number of occurrences based on the Poisson statistics. We employed gaussian fitting function to extract 1 P.E. peak and calculated the absolute gain of PMT. In general, the PMT gain depends on supplied high voltage with following relation:

$$Gain = a^n \left(\frac{V}{n+1}\right)^{kn} = \alpha V^\beta,$$

where $a$ is a constant parameter, the value of $k$ depends on material of electrode, $n$ is the number of dynodes, which is 10 in this case, $\alpha = a^n/(n+1)^{kn}$, and $kn \equiv \beta$. We will operate PMT with a gain of $10^7$. From the requirements of the operation, we required that PMTs have a gain of $10^7$ with high voltage between 1,150 and 1,650 V. Gains for nine different voltages with a step of 50 V were measured to obtain exact HV value for $10^7$ from fitting function.

| Specification | min. | max. |
|---|---|---|
| HV obtaining $10^7$ gain | 1,150 V | 1,650 V |
| P/V ratio | 2.5 | - |
| TTS (FWHM) | - | 4.4 ns |
| Dark count rate | - | 8,000 Hz |
| Skb×2.6 (= QE×CE) | 20.8 % | - |

Table 2: Specification requirements to HPK.

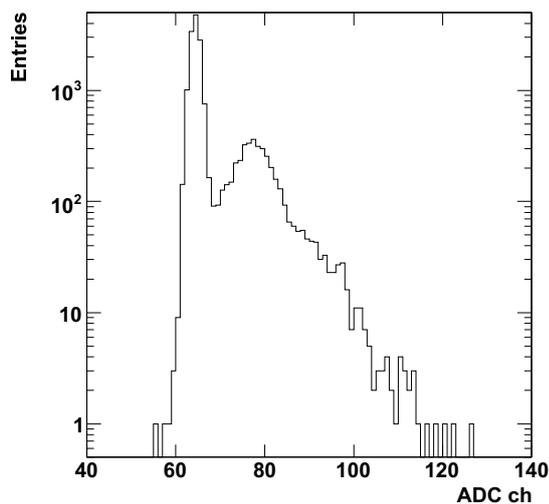

Figure 7: Charge distribution obtained using laser pulse with an intensity of 0.25 P.E. and supplying HV with 1,200 V. The 1 P.E. peak is seen at ADC counts of 80.

Peak to valley (P/V) ratio is an index value for significance of single photo electron signals relative to the noise level or pedestal. It is defined as a ratio of the height of 1 P.E. peak to the valley between the pedestal and 1 P.E. peak. We required P/V ratio to be larger than 2.5 for 0.1 P.E. light injection.

Timing information is crucial to determine the vertex position in the detector. Transit time (TT) is the time difference between the light injection and the signal output. In order to estimate error of the timing of PMT signal, the transit-time spread (TTS) is an important parameter. TTS is defined as the full width at half maximum (FWHM) of the TT distribution. It is affected by the shape of electric field between photo cathode and the first dynode and the structure of dynode in PMT. We required that TTS must be lower than 4.4 ns from the result of simulation study.

Dark count is caused by thermal electrons from photocathode and dynodes, ionization in the PMT and so on. Too much dark count rate would become serious background because only a few photons per PMT are expected in a neutrino event. Therefore, we should check the frequency of dark counts. We required the dark count rate must be lower than 8,000 Hz with 0.25 P.E. threshold after 20 hours of idling.

Quantum efficiency (QE) is conversion probability from photon to photoelectron at PMT cathode, and collection efficiency (CE) is defined as dynode collection efficiency of converted photoelectron. Those two values are difficult to measure separately as both affect the charge output of PMT with same way. We can only measure the product of those values (QE×CE). Different index is used at the inspection by HPK. The value, Skb, means "Cathode blue sensitivity index" and is empirically known that QE×CE (%) on the central region of PMT surface obtained by our measurement is equivalent to Skb×2.6. We required the value of Skb×2.6 must be larger than 20.8 %.

## 3. PMT evaluation systems

As was described in Section 1, we took two steps on evaluation of 400 PMTs delivered to Japan. Hence, two types of evaluation systems had been developed.

First testing after delivery was carried out with the Step-1 evaluation system by which performance of each PMT was individually evaluated in detail. Evaluation subjects are HV extracting $10^7$ gain, peak to valley ratio, time specifications, dark count rate and QE×CE. Especially, the system is designed to measure QE×CE uniformity within photo cathode.

Second testing step after transportation to Germany was carried out by using the Step-2 evaluation system. The main purpose of this test is to check transportation damage by evaluating HV value to give $10^7$ gain, peak to valley ratio and dark count rate, and compare them to those measured in Japan. In order to reveal a possible failure, LED light pulse for 20 hours was applied before this evaluation test as PMT aging operation. In this section, the details of our evaluation systems are described.

### 3.1. Step-1 evaluation system

Figure 8 shows the design of Step-1 evaluation system. The system is set up in a dark box made of aluminum and one can open the front door to set a PMT.

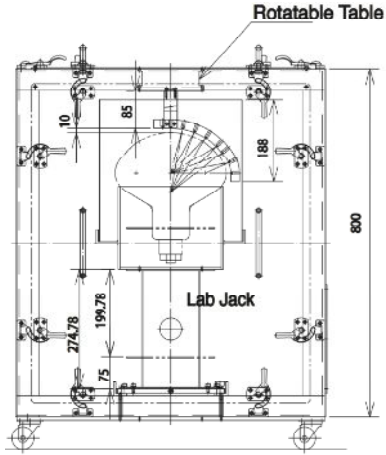

Figure 8: Design of Step-1 evaluation system.

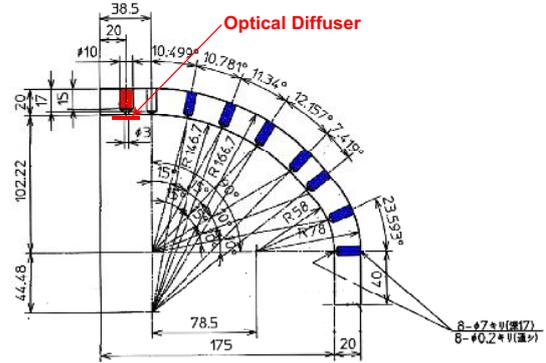

Figure 9: The light arm for the PMT evaluation. The top port is to insert the optical fiber for laser and 7 other ports are to insert LEDs.

Cylindrical $\mu$-metal surrounds the side of the PMT to shield the geo-magnetic field. The size of $\mu$-metal is 315 mm in diameter and 325 mm in height. Its reduction rate of magnetic field was measured to be one-tenth by a handy gauss meter.

Pico-second laser pulser and 7 LEDs are equipped as shown in Figure 9. A pico-second laser pulser, 1 MHz PiLas produced by Advanced Laser Diode Systems GmbH, is used to evaluate PMT with single photoelectron signal. The wavelength is 438.7 nm, which is equivalent to spectrum peak of scintillation light from liquid scintillator for Double Chooz. Time duration is less than 20 ps. A frost type diffuser was attached at the exit of the optical fiber to distribute the light on the surface of PMT. The distance from the light source to the surface of a PMT was set to about 20 cm and all evaluation items except for QE×CE was measured with this setup.

7 LEDs are attached to the arm along the PMT surface. The arm is rotated by stepping motor system and used to measure QE×CE uniformity on the surface of PMT. The mounted LEDs emit light with wavelength of 430 nm. It is designed to incident the light vertically to the photo cathode. The diameter of holes for light from LEDs are 0.2 mm. A manual-adjustable jack can control the vertical position of PMT. When the photon detection efficiency was measured, PMT was lifted up with the jack.

*3.2. Step-2 evaluation system*

Figure 10 shows the design of the system to evaluate 8 PMTs simultaneously. The box is made of aluminum. One can open the front two doors in order to place 8 PMTs and $\mu$-metals. There are four pairs of rails as guide to set the PMT position accurately. Two bars are equipped to route optical fibers for light injection to each PMT.

Only one LED, the same LED as used in the Step-1 evaluation system, was used. The LED is placed outside of the box and the lights are divided into 8 optical fibers with light injector. The light intensity was adjusted with a neutral-density filter placed between LED and fibers. These fibers are routed to upper side of each PMT along the guide bars inside the box. The light is injected to each PMT through the collimator and diffuser. A pulse from NIM module is used as a LED driver after conversion to TTL pulse and a trigger for ADC gate.

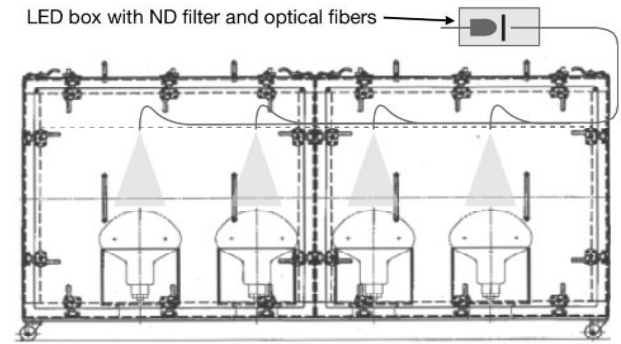

Figure 10: Design of Step-2 system. One can access 4 lanes from two front doors, and set 2 PMT at backward and forward in each line.

4. Evaluation of performance

All PMTs delivered to Tokyo Institute of Technology were evaluated using Step-1 evaluation system by the following procedure:

1. Turn on high voltage supply and wait for 1 hour to stabilize the PMT.
2. Measure the gain with 9 different HVs to define the HV obtaining $10^7$ gain and set the HV for $10^7$ gain.
3. Measure P/V ratio, TT and TTS.
4. Record pulse shape with a digital oscilloscope.
5. Measure dark count rate.
6. Open the box and raise the PMT by jack. Measure the QE×CE map.

Three or four PMTs per day were evaluated.

Step-2 evaluation after transportation to Germany was performed as follows:

1. Turn on the high voltage supply and illuminate PMTs with dozens of photoelectrons for 20 hours as aging operation.
2. Measure the gain supplying 9 different HVs to define the HV obtaining $10^7$ gain. Change the HV for $10^7$ gain if necessary.
3. Measure P/V ratio.
4. Measure dark count rate.

Overall, it took about 6 months to evaluate 400 PMTs in this scheme. The temperature was kept between 20 and 25 degrees with an air-conditioner in both test procedures.

*4.1. Gain*

The gain of PMT is written as:

$$Gain = \frac{2 \times ADC_{count} \times ADC_{gain}}{G_{amp} \times e},$$

where $ADC_{count}$ is the ADC value of single photoelectron peak after subtracting the pedestal value, $ADC_{gain}$ is charge per ADC count, $G_{amp}$ is additional amplifier gain which is used to get more charge to match the range of ADC module and $e$ is electron charge. Multiplication by 2 is needed to take into account the back-termination effect of the base circuit. Figure 11 shows the gain as a function of voltage supplied. The value of HV to obtain $10^7$ gain was calculated by the fitting result and this value was applied for the following measurements.

Figure 12 shows the distributions of high voltage for 390 PMTs which gives $10^7$ gain. The distributions from HPK and our evaluation systems are in good agreement.

*4.2. Peak to valley ratio*

In order to measure P/V ratio, 300,000 events were taken with low light intensity corresponding to 0.1 P.E. of expected number of occurrences based on the Poisson

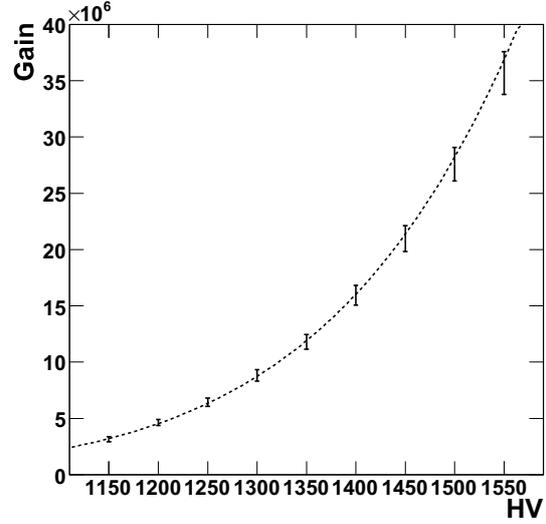

Figure 11: Gain as a function of high voltage. In this example, the HV obtaining $10^7$ gain is 1,320 V.

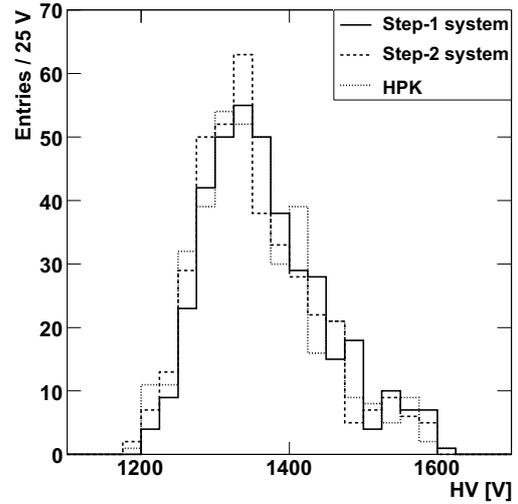

Figure 12: High voltage values for 390 PMTs which give $10^7$ gain. Solid and dashed lines show the results obtained by Step-1 system and Step-2 system, respectively. Dotted line shows specification measured by HPK.

statistics. An *exponential + gaussian* function was used to fit and to calculate the P/V ratio:

$$f(x) = P_0 \times exp\left(-\frac{x}{P_1}\right) + \frac{P_2}{\sqrt{2\pi}P_4} exp\left(-\frac{(x-P_3)^2}{2P_4^2}\right).$$

Number of events at peak and valley was extracted from this fitting function, then converted to the P/V ratio. Figure 13 shows the single photoelectron charge distribution and a result of fitting.

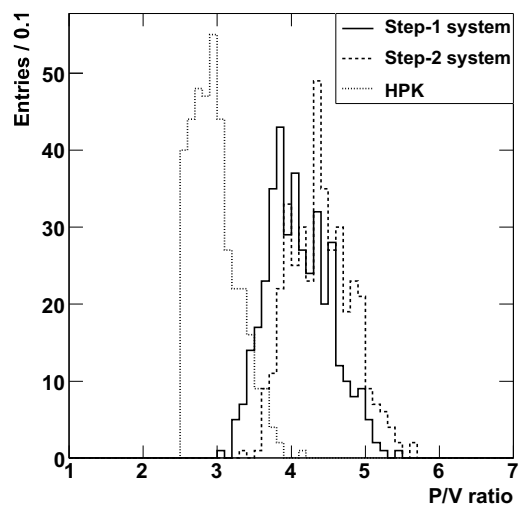

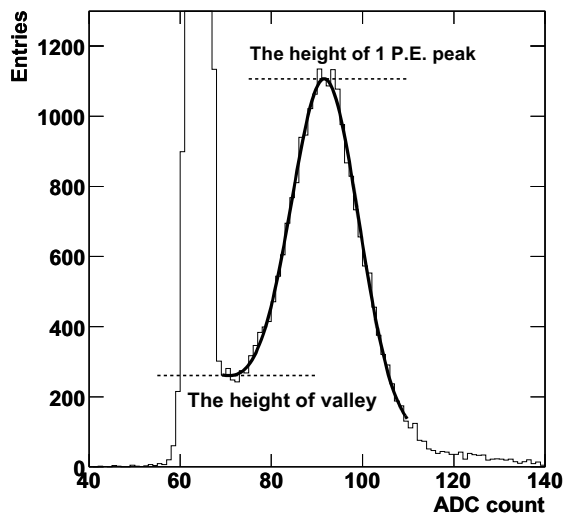

Figure 13: The charge distribution obtained using the laser pulse with an intensity for 0.1 P.E.. The height ratio of 1 P.E. peak to the valley between the pedestal peak and 1 P.E. peak is calculated as "Peak to valley ratio". In this example, P/V ratio is 4.2.

Figure 14: Evaluation result of peak to valley ratio for 390 PMTs. Solid and dashed lines show the results obtained by Step-1 system and Step-2 system, respectively. Dotted line shows the specification measured by HPK.

Result of P/V ratio evaluation is summarized in Figure 14. There is a significant difference between HPK and our measurements. This difference is attributed to different measurement setup, such as different noise-cut capacitor used in splitter circuit, the distance from light source with different diffused angle and use of $\mu$-metal. Our measurement condition is more appropriate to evaluate the PMT performance for the Double Chooz experiment. Left cut off of HPK measurement corresponds to the requirement as shown in Table 2.

### 4.3. Dark count rate

Dark count rate was measured with a threshold level of 0.25 P.E. pulse height. Generally it is known that strong light to PMT surface causes glass phosphorescence and increases count rate for a while [5]. It will attenuates as time goes from turning on high voltage as shown in Figure 15. We measured the time dependence of count rate during 30 minutes, and use the mean as the result.

Result of dark count rate measurement is shown in Figure 16. There are differences among Step-1, Step-2 evaluations and HPK result. HPK measured dark count rate after 30 minutes of turning HV on, while we waited 3 hours in Step-1 test and 20 hours in Step-2 test before the measurement. Therefore, our Step-1 test result is lower than HPK measurement and Step-2 result is further lower than 3 kHz.

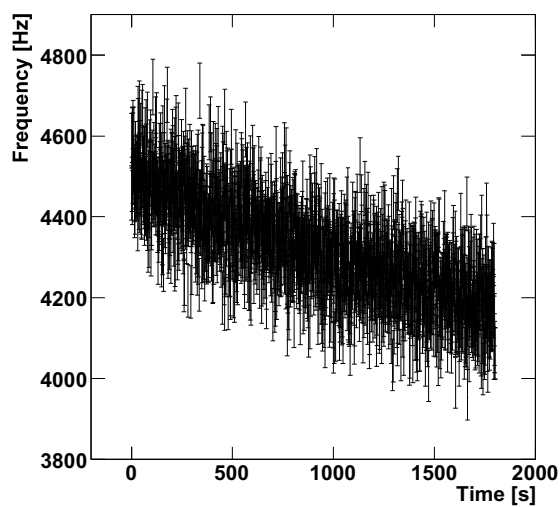

Figure 15: The dependence of the dark count rate on time measured by the Step-1 system. This measurement had started approximately 3 hours after turning HV on.

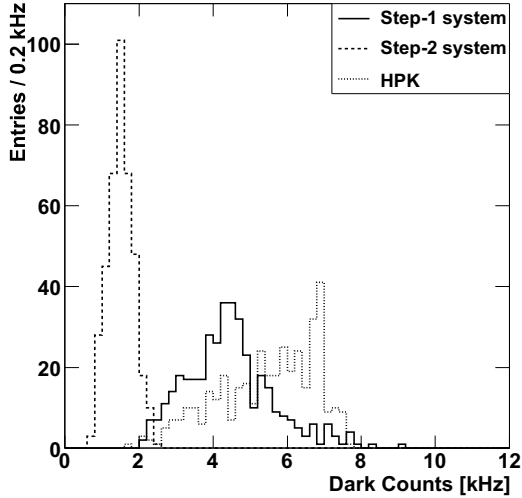

Figure 16: Evaluation result of dark count rate for 390 PMTs. Solid and dashed lines show the results obtained by Step-1 system and Step-2 system, respectively. Dotted line shows the specification measured by HPK.

*4.4. Time specification*

Transit time (TT) and its spread (TTS) are measured with a precision of sub-nano second using the "Picosecond" laser pulse. In the measurement, TDC was started with the trigger signal from the laser pulser and stop signal was created by a discriminator with threshold level of 0.25 P.E. pulse height. The absolute TT could not be measured since the trigger-start time from laser pulse was not defined absolutely but we measured relative difference of TT among 390 PMTs. Time broadening shows asymmetric gaussian-like distribution as shown in Figure 17. Hence, asymmetric gaussian was employed as a fitting function to extract relative TT and TTS. TTS is defined as full width at half maximum (FWHM) in the distribution of time broadening.

The result of TTS is summarized in Figure 18. The result of our measurement is better than HPK result. It may be explained by the setup difference such as the length between light source and PMT surface since TTS contains the position dependence of TT on PMT surface. However, a good correlation was observed between the two measurements.

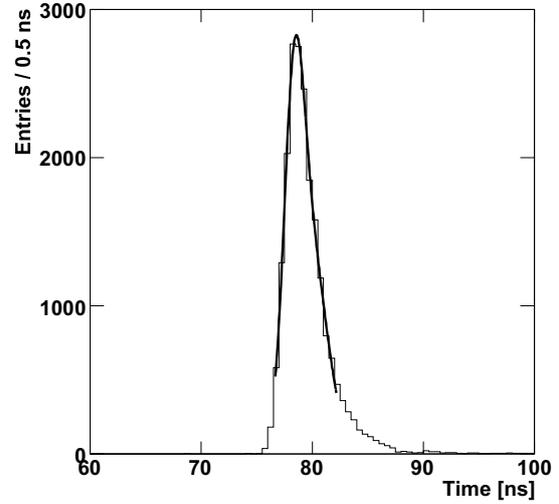

Figure 17: Time broadening distribution obtained by shining the laser pulse with an intensity for 0.1 P.E.. The peak and FWHM are used for relative TT and TTS.

Evaluation result of relative TT is summarized in Figure 19. It was measured by Step-1 system only. In order to validate the results, the high voltage dependence of the relative TT was checked. It is known that there is a strong correlation due to different velocity of electron in different electric field between cathode and first dynode, and strong correlation was confirmed as shown in Figure 20.

*4.5. QE×CE and its uniformity*

QE×CE value is calculated by the ADC distribution for each position on PMT surface with the following function:

$$\mathrm{QE} \times \mathrm{CE} = \frac{N_{pe}}{N_p} = \frac{2 \times ADC_{count} \times ADC_{gain}}{N_p \times e \times Gain},$$

where $N_{pe}$ is the number of measured photoelectrons. $N_p$ is the number of injected photons calibrated with a reference 10-inch PMT before each measurement of QE×CE value because it can fluctuate. QE × CE value includes non-uniformity effect of gain since there is contribution of gain in this relation. The average $N_{pe}$ was 200 photoelectrons intensity, which is small enough comparing with a dynamic range of this PMT. Figure 21 shows the evaluation result of a PMT having 135 degrees of the first dynode direction against cable output (+x axis).

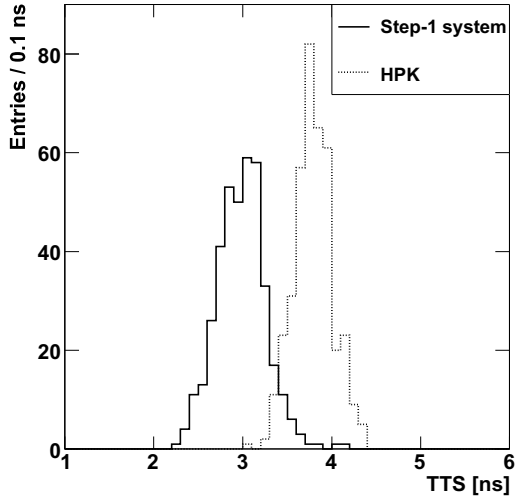

Figure 18: Evaluation result of TTS for 390 PMTs. Solid line shows the results obtained by Step-1 system. Dotted line shows the specification measured by HPK.

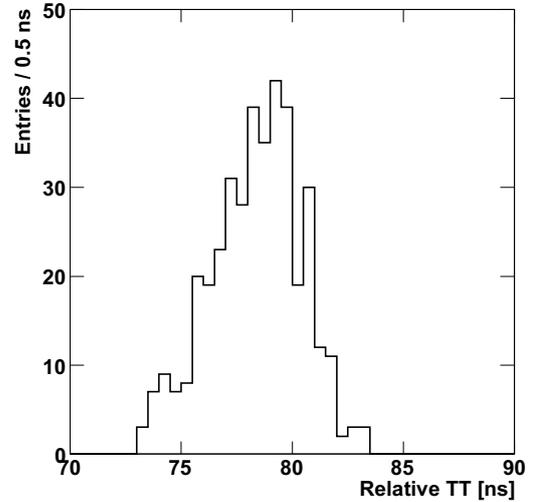

Figure 19: Evaluation result of relative TT for 390 PMTs. Solid line shows the results obtained by Step-1 system.

The result of average QE×CE is summarized in Figure 22. The mean QE×CE values were calculated from 72 measurement points on the inner 3 rings as a average QE×CE of PMT. QE×CE uniformities were investigated individually as shown in Figure 21. The QE×CE distribution depends on the dynode direction. This non-uniformity of QE×CE was known before detector is designed, so that dynode directions are randomized with 45 degrees unit in order to suppress detecter bias. In addition, typical non-uniformity obtained by this measurement is used for MC input for more precise PMT characterization.

*4.6. Results of PMT evaluation*

Finally, 400 PMTs were successfully evaluated. A few PMTs were rejected by our evaluation with Step-1 system. However, HPK replaced the defective products. 390 PMTs were selected as installation candidates in consideration of the requirements in Table 2. Evaluation results of those selected PMTs are summarized in Table 3 and Table 4. Results of HPK, Step-1 and Step-2 measurement are consistent. The attribution of different distributions in Figure 14, 16 and 18 are understood. No difference is found in the other distributions. Remaining 10 PMTs are kept as spare ones. Those PMTs show relatively low QE×CE, unstable dark count rate or large non-uniformity of gain over the position of photocathode. Those, however, can be installed to detector since variations of performance are considered within our requirements.

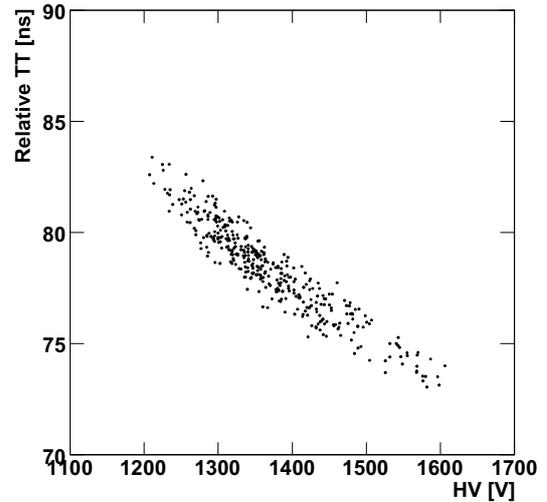

Figure 20: Relative TT obtained by Step-1 system as a function of applied HV. The dependence is roughly 2.5 ns/100V

## 5. Summary

The aim of the Double Chooz experiment is to accurately measure the neutrino mixing angle $\theta_{13}$ with anti-electron neutrinos from nuclear reactors. Very low-background 10-inch PMTs were developed in order to achieve this very precise measurement. PMT evaluation was carried out with two types of evaluation system after delivery and transportation. The performances for each PMT were obtained in detail and no failure was found during transportation and aging operation. As a result, 400 PMTs including 10 spares which fulfill our requirement have been prepared successfully. Those result are stored to database, and will be used in initial calibration values, analysis and MC simulation.

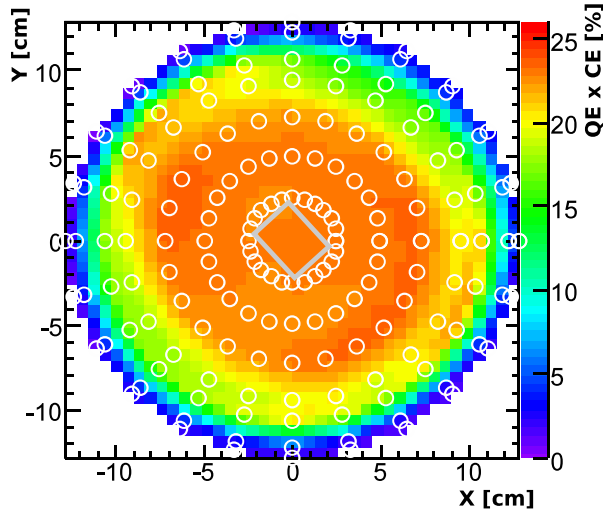

Figure 21: Result of QE×CE for a typical PMT. This PMT has 135 degrees of the first dynode direction with respect to cable output. Cable output is along the x-axis. The box at the center indicates the direction of the first dynode of the PMT. White circles show the measurement points.

| Specification | min. | max. |
| --- | --- | --- |
| HV obtaining $10^7$ gain | 1,210 V | 1,610 V |
| P/V ratio | 3.0 | 5.5 |
| TTS (FWHM) | 2.3 ns | 4.1 ns |
| Dark count rate | 2,100 Hz | 9,000 Hz |
| Skb×2.6 (= QE×CE) | 20.3 % | 26.7 % |

Table 3: Result of Step-1 evaluation system.

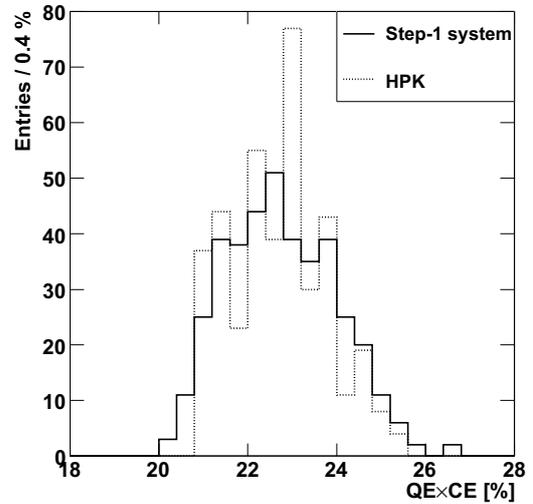

Figure 22: Evaluation result of average QE×CE for 390 PMTs. Solid and dotted lines show the results obtained by Step-1 system and HPK. The result of Step-1 system show the mean of three injected positions from center and whole angles. HPK result is Skb×2.6. Skb means blue sensitivity as described in the Section 2.

| Specification | min. | max. |
| --- | --- | --- |
| HV obtaining $10^7$ gain | 1,190 V | 1,590 V |
| P/V ratio | 3.4 | 5.6 |
| Dark count rate | 700 Hz | 2,600 Hz |

Table 4: Result of Step-2 evaluation system.

## Acknowledgement

This work was performed under a support by Grant-in-Aid for Specially promoted Research, funded by Ministry of Education, Culture, Sport and Technology of Japan (MEXT).

We appreciate Dr. N. Bowden and Dr. A. Smith for measuring glass radioactivity. Double Chooz PMT group consists of Japan, Germany, Spain and US collaborators. We also thank our collaborators for their significant contribution to realize the Double Chooz PMT system.